\documentclass[pre,twocolumn,a4paper,showpacs]{revtex4-1}
\usepackage{amssymb}
\usepackage{wasysym}
\usepackage{graphicx}
\usepackage{epsfig}
\usepackage{subfigure}
\begin{document}

\title{Effect of group organization on the performance of cooperative processes}

\author{Sandro M. Reia and Jos\'e F. Fontanari}
\affiliation{Instituto de F\'{\i}sica de S\~ao Carlos,
  Universidade de S\~ao Paulo,
  Caixa Postal 369, 13560-970 S\~ao Carlos, S\~ao Paulo, Brazil}
  

\begin{abstract}
Problem-solving competence at  group level is influenced by the structure of the  social networks and so it may shed light on  the   organization  patterns of gregarious animals. Here we use an agent-based model to investigate whether the ubiquity of  hierarchical networks in nature  could be explained as the result of a selection pressure favoring problem-solving efficiency. The task of the agents  is to find the global maxima of  NK fitness landscapes and  the agents cooperate by broadcasting  messages  informing on their fitness to the group.
This information is then used to imitate, with a certain probability,  the fittest agent in their influence networks. For rugged landscapes, we find that the modular organization of the hierarchical network with its  high degree of clustering eases the escape from the  local maxima, resulting in a superior performance as compared with the  scale-free and the random networks. The optimal performance in a rugged landscape is achieved by letting the main hub to be only slightly more propense to imitate the other agents than vice versa. The performance is greatly harmed  when the main hub carries out the search independently of the  rest of the group  as well as when it compulsively imitates  the other agents.  
\end{abstract}

\maketitle

\section{Introduction}

There is little dispute over the claim that the collective structures built by termites, ants and slime molds are  products of cooperative work  performed by a myriad of  organisms who, individually, are  inept to  conceive  the  greatness of the structures they build \cite{Marais_37}.   Thinking of   those collective structures as the organisms'  solutions to the problems that endanger their existence, it is natural to argue that competence in problem solving should be viewed as a candidate  selection pressure  for molding the  organization of groups of social animals \cite{Bloom_01,Queller_09}.  
 
Information flows between individuals via social contacts and, in the problem-solving  context, the relevant process is imitative learning as  expressed in this quote by Bloom  ``Imitative learning acts like a synapse, allowing information to leap the gap from one creature to another'' which summarizes his  view of those collective structures as global brains  \cite{Bloom_01}. Evidences that cooperative work powered by social learning is an efficient process to solve difficult problems are  offered by the    variety of social learning based  optimization heuristics, such as the particle swarm optimization algorithm \cite{Bonabeau_99} and the adaptive culture heuristic \cite{Kennedy_98,Fontanari_10}.

From the perspective of the computer science, there has been considerable progress on the  understanding of  the factors that make cooperative group work effective \cite{Clearwater_91,Clearwater_92, Page_07}, although, somewhat disturbingly, the most popular account of 
collective intelligence, the so-called wisdom of crowds, involves the suppression of cooperation  since  its success depends on the individuals making their guesses independently of each other \cite{Surowiecki_04} (see, however, \cite{King_12}).

In this contribution we build on a recently proposed  minimal model of distributed cooperative problem-solving systems  based on imitative learning \cite{Fontanari_14} to study the influence of the social network topology on the performance of cooperative processes. Individuals cooperate by broadcasting  messages  informing on their fitness and use this information to imitate, with a certain probability, the fittest individual in their influence networks. The task of the individuals is to find the global maxima of smooth and rugged fitness landscapes generated by the NK model \cite{Kauffman_87,Kauffman_89} and the performance or efficiency of the group is measured by the number of trials required to find those maxima.  

Our goal is to investigate whether the ubiquity of hierarchical networks  \cite{Ravasz_03}, which are both modular and scale-free,  could be explained as the result of a selection pressure favoring problem-solving efficiency. In fact, for rugged landscapes we find that the hierarchical network  performs better 
than  the scale-free and the random networks. The modular organization of the hierarchical network with its  high degree of clustering facilitates the system to escape  local maxima, despite the presence  of a hub with very high connectivity (super-spreader) which, in general, may cause great  harm to  the system performance by broadcasting misleading information about the location of the global maxima \cite{Francisco_16} as happens in the case of the scale-free network. For smooth landscapes, the topology of the network has little influence on the performance of the imitative search.

In addition, we find that for the three network    topologies considered here, namely, hierarchical, scale-free and random topologies, allowing the main hub (i.e., the node with the highest degree) to explore the landscape without much consideration for the other individuals, even though those individuals may learn from  it, is always detrimental to the performance of the system.  Interestingly,  for the hierarchical and scale-free networks, the optimal performance in a rugged landscape is achieved by letting the main hub to be only slightly more propense to imitate the other agents than vice versa.  A compulsive imitator located at  the main hub of the hierarchical network (or at the two main hubs of a scale-free network) leads to a disastrous performance. For the random network, where the main hub is not very influential, the performance is maximized by the compulsive imitation strategy. This  is also true  for the three topologies  in the case of a smooth landscape, but the reason is that in the absence of local maxima  it is always better to imitate  the fittest individual in the group.

The rest of this paper is organized as follows.  For the sake of completeness,  we present a brief description of the NK model of  
rugged fitness landscapes  in Section \ref{sec:NK}. The rules of the agent-based model that implements the imitative search are explained in Section \ref{sec:imit} and the three network topologies -- hierarchical, scale-free and random -- are presented in Section \ref{sec:net}. 
In  Section \ref{sec:res} we present and discuss the results of the simulations of the imitative search on rugged and smooth NK landscapes
for those three topologies.  Finally, Section \ref{sec:disc} is reserved to our concluding remarks.

\section{NK Model of Rugged Fitness Landscapes}\label{sec:NK}

The NK model  is a computational framework  to generate  families of statistically identical rugged fitness landscapes.   It  was  proposed by Stuart Kauffman in the late 1980s aiming at  modeling  evolution  as an incremental process, the so-called adaptive walk, on rugged landscapes \cite{Kauffman_87,Kauffman_89}. Today the NK model is the paradigm of problem spaces with many local optima, being  particularly popular  among the  organizational and management research community \cite{Levinthal_97,Lazer_07,Fontanari_16}.

The NK  model  is named for the two integer parameters that are used to randomly generate  landscapes, namely, $N$ and $K$. The landscape  is defined in the space of binary strings of length $N$ and so this parameter determines the size of the solution space, $2^N$.  The other parameter  $K =0, \ldots, N-1$ influences the ruggedness of  the landscape. In particular, the correlation between the fitness of any two neighboring strings (i.e., strings that differ at a single component) is $ 1 - \left ( K+1 \right )/N $ \cite{Kauffman_89}. Hence $K=0$ corresponds to a smooth landscape whereas $K=N-1$  corresponds to a completely  uncorrelated landscape.
For concreteness, next  we describe briefly the procedure to generate a random realization of a NK landscape.

The $2^N$ distinct binary strings of length $N$ are  denoted by $\mathbf{x} = \left ( x_1, x_2, \ldots,x_N \right )$ with
$x_i = 0,1$. To each string $\mathbf{x}$ we associate a fitness value $\Phi \left ( \mathbf{x}  \right ) $ which  is an average  of the contributions from each 
component $i$ in the string, i.e., $\Phi \left ( \mathbf{x}  \right ) =  \sum_{i=1}^N \phi_i \left (  \mathbf{x}  \right )/N$,
where $ \phi_i$ is the contribution of component $i$ to the  fitness of string $ \mathbf{x} $. It is assumed that $ \phi_i$ depends on the state $x_i$  as well as on the states of the $K$ right neighbors of $i$, i.e., $\phi_i = \phi_i \left ( x_i, x_{i+1}, \ldots, x_{i+K} \right )$ with the arithmetic in the subscripts done modulo $N$.  The functions $\phi_i$ are $N$ distinct real-valued functions on $\left \{ 0,1 \right \}^{K+1}$ but
 the usual procedure is to assign to each $ \phi_i$ a uniformly distributed random number  in the unit interval \cite{Kauffman_89}, which then guarantees that $\Phi \in \left ( 0, 1 \right )$ has a unique global maximum. 
 
 For $K=0$ the global maximum is the sole maximum of $\Phi$, which can  be easily found by picking for each component $i$ the state $x_i = 0$ if  $\phi_i \left ( 0 \right ) >  \phi_i \left ( 1 \right )$ or the state  $x_i = 1$, otherwise. For $K=N-1$, the (uncorrelated) landscape  has on the average  $2^N/\left ( N + 1 \right)$ maxima with respect to single bit flips \cite{Derrida_81}. Finding   the global maximum of the NK model for $K>0$ is a NP-complete problem \cite{Solow_00}, which  means that the time required to solve the problem using any currently known deterministic algorithm increases exponentially fast with the length $N$ of the strings \cite{Garey_79}.

We note that the specific features of a realization of the NK landscape (e.g., number and location of the local maxima with respect to the global maximum) are not fixed by the parameters $N$ and $K$, because the components $ \phi_i$ are chosen randomly in the unit interval. This is the reason that  finding the global maximum for {\it any} realization of the NK landscape for large $N$ and  $K>0$ is an extremely difficult computational problem \cite{Solow_00}.  Hence,  in order to better apprehend  the  influence of the network topology and, in particular, 
 the role of the main hub  on the  performance  of  cooperative problem-solving systems, 
 here we use a single realization of the NK fitness landscape for fixed values of $N$ and $K$.

More pointedly, we consider two types of landscape: a smooth landscape with  $N=16$ and $K=0$  and  a rugged landscape with $N=16$ and $K=5$.  Since for $K=0$  all NK landscapes are equivalent, there is no lack of generality in considering a single instance of that family.
The particular realization of the NK landscape with  $N=16$ and $K=5$ considered here exhibits 296  maxima in total, among which 295 are local maxima.  The mean relative fitness  of the local maxima  with respect to the fitness of the global maximum is $0.81$ whereas the  mean relative  fitness of all strings is $0.60$. It is interesting to  note that for large $N$  the NK model exhibits the so-called  complexity  catastrophe \cite{Kauffman_89}, i.e., as $N$ increases the  fitness of the local  maxima become poorer  to  such  a point  that  they  are  not  better  than  the fitness of a  randomly chosen  string.  
The effects of averaging over different  realizations of the rugged landscape is addressed briefly  at the end of Section \ref{sec:res}.

\section{Imitative learning search}\label{sec:imit}

We consider a  system composed of $L$  agents $i=1,\ldots,L$ and assume that  each agent operates in an initial binary string drawn at  random with equal probability for the bits $0$ and $1$. Agent $i$  can choose  between two distinct processes to operate  on its string. 
The first process,
which happens with probability $1-p_i$,  is the elementary move in the solution space that consists of picking a  bit  at random   from  the string and flipping it. This elementary move allows the agents to explore in an incremental way the entire solution space formed by the $2^N$  binary strings. The second process, which  happens with probability $p_i$, is the  imitation  of a model string. Here the model string is defined as  the string that exhibits the largest fitness value  among the (fixed) subgroup of agents that  can influence (i.e., are connected to)  agent $i$. The  model string and the string $i$ (i.e.,  the string operated by agent $i$)  are compared  and the different bits are singled out.  Then agent  $i$ selects at random one of the distinct bits and flips it so that this bit is now the same in both string $i$ and the model  string. After imitation these two strings become more similar, as expected. In the case the  string $i$ is identical to the model string,   agent $i$  executes the elementary move with probability one.   

The parameter $p_i \in \left [0,1 \right ]$ is the imitation propensity of agent $i$. The case $p_i=0$ corresponds to the baseline situation in which   agent $i$ explores the solution space independently of the other agents.  
The imitation procedure described above was based on the incremental assimilation mechanism used to study the influence of an external media  \cite{Shibanai _01,Peres_11} in  the celebrated Axelrod's model  of culture dissemination  \cite{Axelrod_97}. We note that an alternative non-incremental imitation procedure, which  allows  string $i$ to  become identical to the model string   by changing  many bits  simultaneously,  may  permanently stuck the search in the  local maxima \cite{Lazer_07}.

The evolution of the system of $L$ agents  proceeds as follows. At each time $t$ we pick an agent at random, say agent $i$, and allow it to operate on its associated string either by imitating the model string or by flipping a bit at random.  Since this operation always results in a change of fitness of string $i$, which may become larger than the fitness of the model string, we need to update the   model string status
at each time.  As usual in such asynchronous update scheme, we choose the time unit as $\Delta t = 1/L$ so that during the increment from $t$ to $t+1$, exactly  $L$ string operations are performed, though not necessarily by $L$ distinct agents.  

The  search ends when one of the agents finds the global maximum and we denote by $t^*$ the halting time. The efficiency of the  search is defined as the total number of string operations necessary  to find the global maximum (i.e., $Lt^*$) 
and so the  computational cost of the search can be defined as $C \equiv L t^*/2^N$, where for convenience we have rescaled $t^*$ by the size of the solution space $2^N$.

In the case of the independent search (i.e., $p_i = 0, \forall i$) the ruggedness of the landscape has no effect on the efficiency  of the search, which depends only on the length of the strings, $N$ and on the system size $L$.  It can be shown that the mean computational cost is given by \cite{Fontanari_15}
\begin{equation}\label{Cind}
\langle C \rangle = \frac{L}{ 2^N \left [ 1 - \left ( \lambda_N \right)^L \right ]},
\end{equation}
where  $\lambda_N$ is the second largest eigenvalue of a tridiagonal stochastic matrix $\mathbf{T}$  with elements  
$T_{ij} = \left ( 1-j/N \right ) \delta_{i, j+1} + j/N \delta_{i, j-1}$ for $j=1,\ldots,N-1$, $T_{i0} = \delta_{i,1}$, and  $T_{iN} = \delta_{i,N}$. Note that
$i=N$ is the only absorbing state of the stochastic process defined by $\mathbf{T}$.
Here $\delta_{i, j} $ is the Kronecker delta and  the  notation $\langle \ldots \rangle$  stands for the average over independent searches on the same landscape. In particular, for $N=16$ we find  $\lambda_{16} \approx 0.9999859$ and since 
$\left ( \lambda_{16} \right)^L \approx e^{- L \left ( 1 - \lambda_{16} \right )}$ we have 
$\langle C \rangle \approx  1.08$ for $L \ll 1/\left ( 1 - \lambda_{16} \right ) \approx 70740$  and
$\langle C \rangle \approx L/ 2^{16} $ for $L \gg 70740$. The first regime, characterized by a  mean computational cost that is independent of the system size $L$, corresponds to the situation where the halting time $t^*$ decreases linearly with increasing $L$. The second regime, where $\langle C \rangle$  increases linearly with $L$, corresponds to the situation where the halting time is $t^* \approx 1$, i.e., the system size is so large that the global maximum is likely to be found already  during the  initial stage when  the strings are generated randomly.

\section{Complex networks}\label{sec:net}

Most real-world  social networks  are scale-free and exhibit  a high degree of clustering, which seems to be independent of the number of nodes $L$ \cite{Albert_02}.  The  scale-free property  means that the probability that a randomly selected node has  degree $k$ obeys a power law  $P \left ( k \right ) \sim k^{-\gamma}$  where the degree exponent $\gamma$  usually varies   between $1$ and $3$. The high degree of clustering is consequence of the  formation of  cliques,  which represent  groups   of  individuals in which every member knows every other member.  Typical scale-free networks produced by the Barab\'asi-Albert algorithm and by its many variants \cite{Barabasi_99} do not exhibit a  high degree of clustering and therefore do not account for the presence of both  properties  of  real networks. The reason seems to be the absence of a modular organization in scale-free networks \cite{Ravasz_03}. To  produce a network that exhibits both modularity (hence, a high degree of clustering)  and the scale-free topology it is necessary to organize the modules hierarchically,   producing the so-called  hierarchical network \cite{Ravasz_03}.

We note, for the sake of completeness, that the clustering coefficient for a node $i$ with $k_i$ links is defined as ${\mathcal{C}}_i =2 n_i /\left [ k_i \left ( k_i-1 \right ) \right ]$, where $n_i$ is the number of links between the $k_i$ neighbors of node  $i$, and the overall level of clustering in a network  can be obtained by averaging those coefficients over all nodes, 
 $\bar{\mathcal{C}} = \sum_{i}^L {\mathcal{C}}_i/L$ \cite{Watts_98}.
 
Here we consider three types of networks, which share the same number of nodes and  links,  but exhibit  very different topologies, as reflected by their degree distributions and their average clustering coefficients. In particular, we consider  hierarchical networks, 
scale-free networks   and random networks. In what follows we describe succinctly each of these networks.
 
\begin{figure}[!h]
\centering\includegraphics[width=0.6\linewidth]{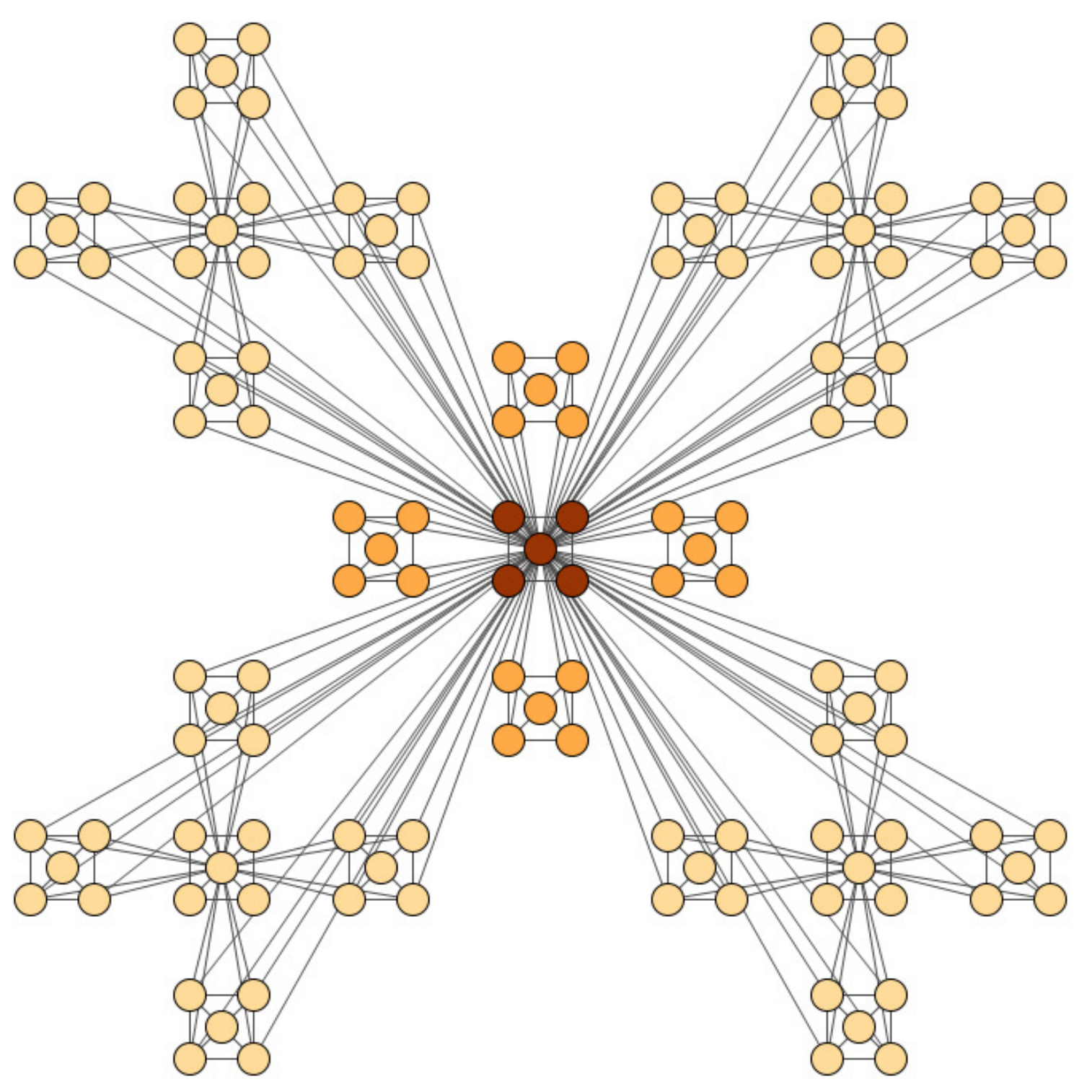}
\caption{(Color online) Hierarchical network with $L=125$ nodes and $394$ links, where each elementary cluster of five nodes is fully connected (the links connecting the diagonal nodes in the elementary clusters are not visible). The   central node of the original cluster at level $1$ has degree $k=84$ and the four central nodes at level $3$ have degree $k=20$. The other nodes have degrees $k=4,5$ and $6$.
The mean clustering coefficient is $\bar{\mathcal{C}} = 0.850$. }
\label{fig:01}
\end{figure}

\subsection{Hierarchical network model}

The starting point of the procedure to construct a hierarchical network that is both modular and scale-free \cite{Ravasz_03} is a cluster of five fully connected nodes, arranged in the corners and in the center of a square. This is level $1$ of the hierarchical network.
Next, four replicas of this square are generated and the nodes at the corners of those replicas  are linked to the central node, resulting in level $2$ of the hierarchy, composed of 20 nodes. Level $3$ is formed by generating four replicas of this 25 nodes module (i.e., the combination of levels $1$ and $2$) and linking the 16 peripheral nodes of each replica to the central node of the old module. Levels $1$, $2$ and $3$ form a new module with 125 nodes. The next step (level $4$) would be to produce again four replicas of this module and connect the peripheral sites to the central node of the old module. If a hierarchical network has $n$ levels then the total number of nodes is $L = 5^{n}$ and the total number of links is $ 6 \left( 5^{n-1} \right)+ 4 \left( 5^{n} - 4^{n} \right)$. Figure \ref{fig:01} illustrates the hierarchical network with $n=3$ levels. There are 40 nodes that exhibit the maximum value of the clustering coefficient, i.e. ${\mathcal{C}}_i=1$ and the main hub at the center of the original module is the node that exhibits the lowest value of this coefficient,
${\mathcal{C}}_i \approx 0.036$.

\begin{figure}[!h]
\centering\includegraphics[width=0.6\linewidth]{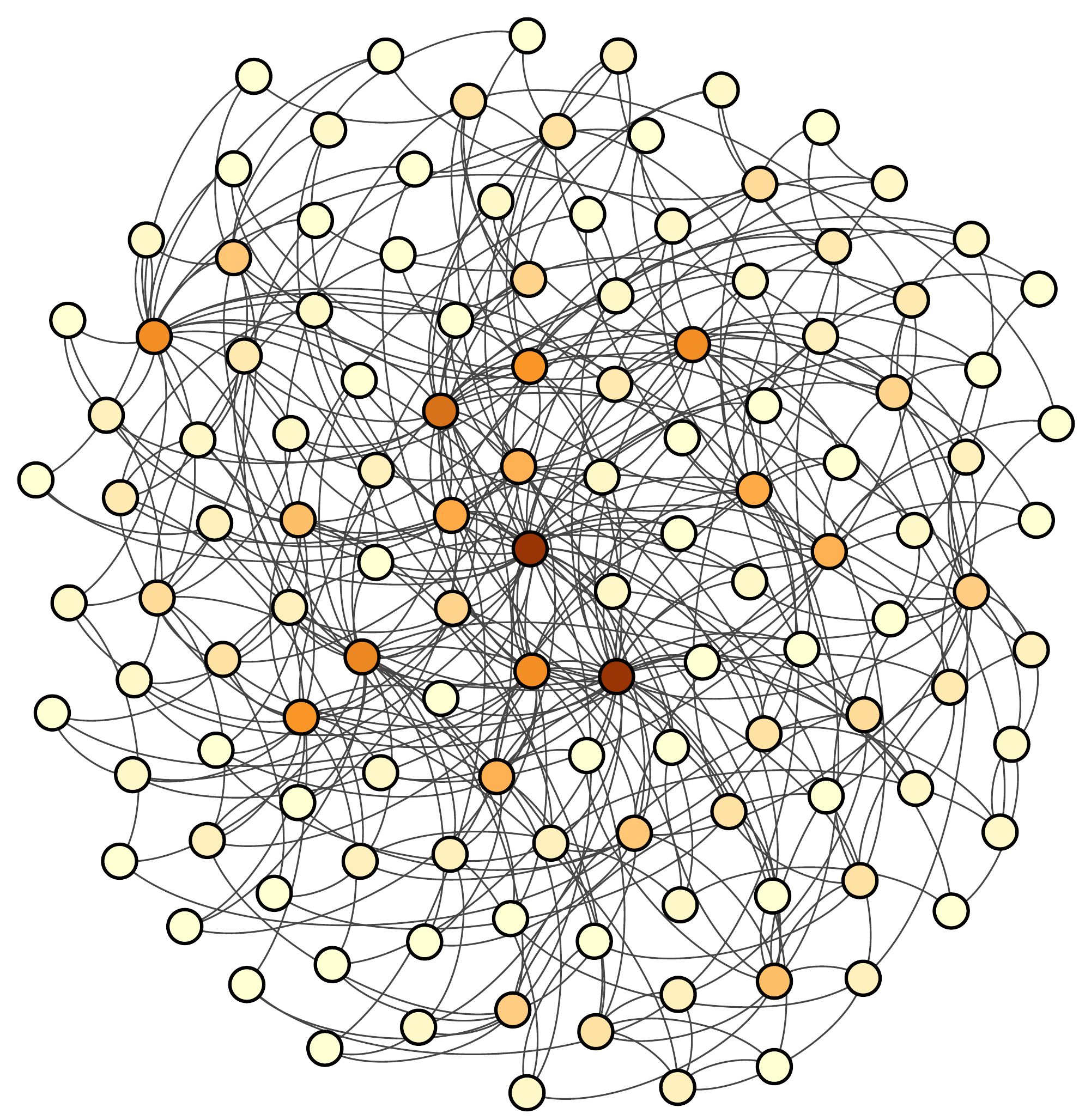}
\caption{(Color online)  Realization of a scale-free network with $L=125$ nodes and $394$ links generated using  preferential attachment and network growth following the Barab\'asi-Albert algorithm  \cite{Barabasi_99}. The highest degree is $k=32$  (2 nodes) and the lowest is $k=3$ (45 nodes). The mean clustering coefficient is $\bar{\mathcal{C}} = 0.180$.}
\label{fig:02}
\end{figure}

%

\subsection{Scale-free network}

A scale-free network is a network whose degree distribution follows a power law when the number of nodes $L$ is very large \cite{Albert_02}, and so the network may exhibit a few nodes  with  very large degrees. 
As already pointed out,  the interesting feature of this  topology is that, while exhibiting the scale-free property, it lacks  modularity which allows us then to examine the influence of this property on the performance of  distributed cooperative problem-solving systems. In order to  generate scale-free networks with a fixed number of nodes (say, 125 nodes) and links (say, 394 links) we have made a minor change in  the classical  Barab\'asi-Albert algorithm \cite{Barabasi_99}. As it is well-known, this algorithm is based on a preferential attachment mechanism and network growth. Beginning 
with $m_0$ disconnected nodes, at each time step, a new node $i$ with $m_0$ links is added to the network. The probability that a node $j$, which is present in the network, will receive a link from $i$ is proportional to the degree of node $j$, i.e., $P(i\to j) = k_j/\sum_l k_l$.  

To generate a scale-free network with a fixed number of nodes and links, we use the following procedure. Given the desired  ratio $r^*$ between the number of  links and the number of nodes, which  in the case of interest is $r^* = 394/125 =3.152$,   we begin the network growing procedure at generation $\tau = 0$ with  $m_0 = \lceil r^* \rceil = 4$ disconnected nodes, so that the ratio at this initial stage is  $r_0 = 0/4 = 0$.  Since $r_0 < r^*$, we add a new node with $m_0$ links  to the original nodes so that at generation $\tau =1$ we have  $r_1 = 4/5$. We keep doing this till growing generation $\tau =15$  at which
$r_{15} = 60/19 > r^*$, when  we add then a new node with  $m_0 -1 = 3$ links, yielding $r_{16} = 63/20 < r^*$ at generation $\tau=16$. The next node which will form generation $\tau=17$ will have then $m_0$ links. The idea  is that if at growing generation $\tau -1$ we have $r_{\tau-1} < r^*$ the new node at generation $\tau$  will have $m_0$ links, otherwise it will have $m_0-1$ links. The network is complete when the number of nodes reaches the desired value, $L=125$ in our case. In Fig.\ \ref{fig:02} we exhibit a realization of a scale-free network
produced by the procedure just described. There are $32$ nodes with ${\mathcal{C}}_i=0$, (i.e., there are  no links between the nodes that are linked to node $i$) and the highest clustering coefficient is ${\mathcal{C}}_i=0.67$.

\subsection{Random network}

\begin{figure}[!h]
\centering\includegraphics[width=0.6\linewidth]{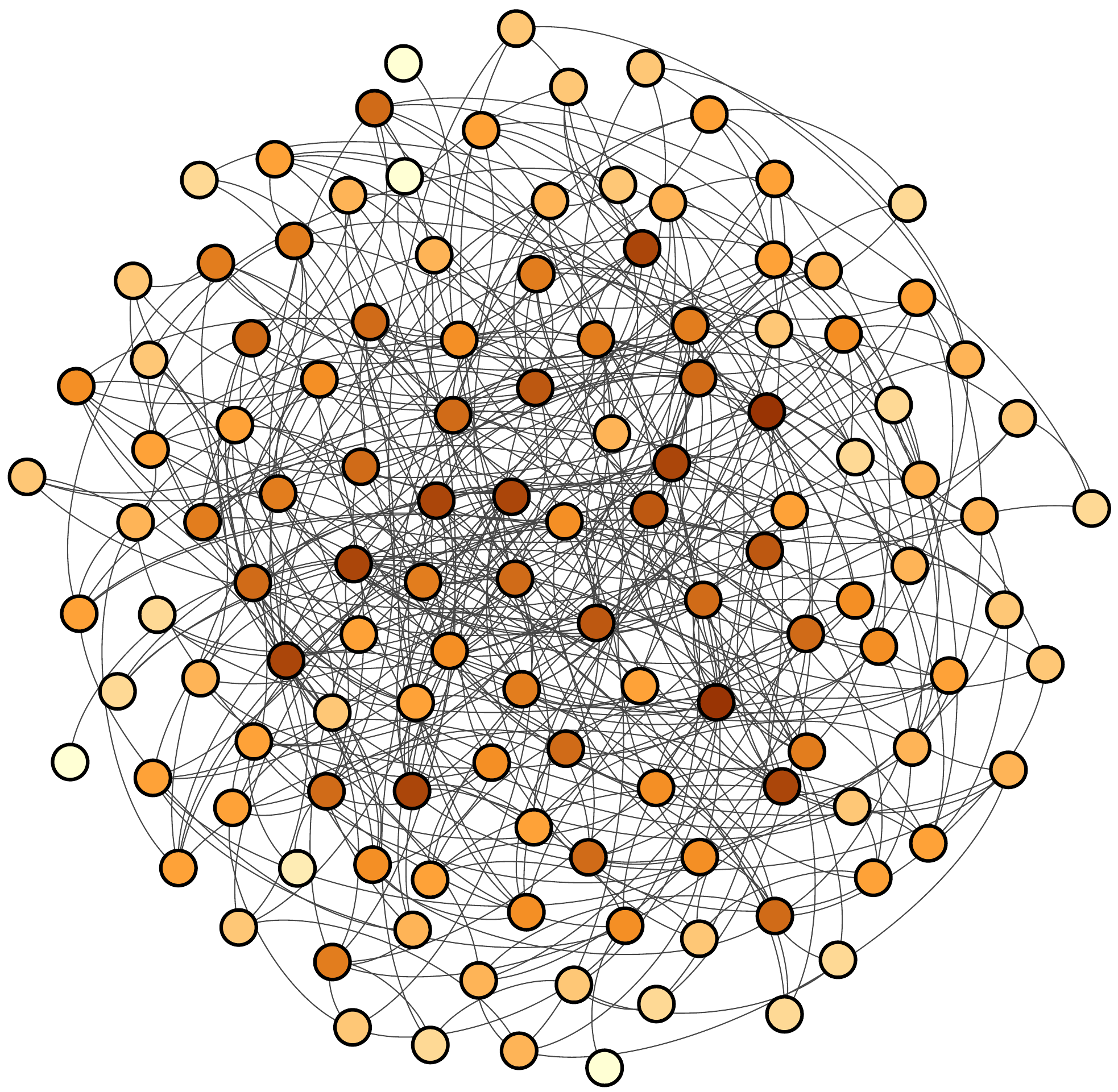}
\caption{(Color online)  Realization of a random network with $L=125$ nodes and $394$ links. The highest degree is $k=12$  (2 nodes) and the lowest is $k=1$ (4 nodes). The mean clustering coefficient is $\bar{\mathcal{C}} = 0.057$.}
\label{fig:03}
\end{figure}

The random network offers a good baseline for  comparison with the two structured networks described before. In this case we simply distribute randomly and without replacement the fixed number of links (say, 394) among the $L\left ( L-1 \right)/2$ pairs of nodes. Figure \ref{fig:03} exhibits a realization of a random network. There are  $50$ nodes with ${\mathcal{C}}_i=0$ and the highest clustering coefficient is ${\mathcal{C}}_i=0.33$.

Finally, we note that while the procedure to construct the hierarchical network is deterministic (i.e., the resulting network is unique), the procedures to generate the scale-free and the random networks are stochastic and so each time those procedures are implemented a different network is produced. Figures \ref{fig:02} and \ref{fig:03} then illustrate typical realizations of those two network topologies.

\section{Results}\label{sec:res}

We begin this section  with the analysis of the performance of the imitative search on a rugged NK  landscape with parameters $N=16$ and $K=5$ (see
Section \ref{sec:NK}). To better appreciate the effects of the local maxima on the performance of the  search we
conclude the section with the analysis of a smooth landscape with parameters $N=16$ and $K=0$.  Although the topology of the networks connecting those agents is variable (i.e., hierarchical, scale-free and random),  the average connectivity of the networks as well as the number of nodes are fixed.  The mean  computational cost is calculated by averaging the computational cost over $10^5$ distinct searches. For the scale-free and  the random networks, for each of those searches we generate a different network. In all figures exhibited in this section, the error bars are smaller than the symbol sizes.

\subsection{Rugged Landscape}

\begin{figure}[!h]
\centering\includegraphics[width=0.9\linewidth]{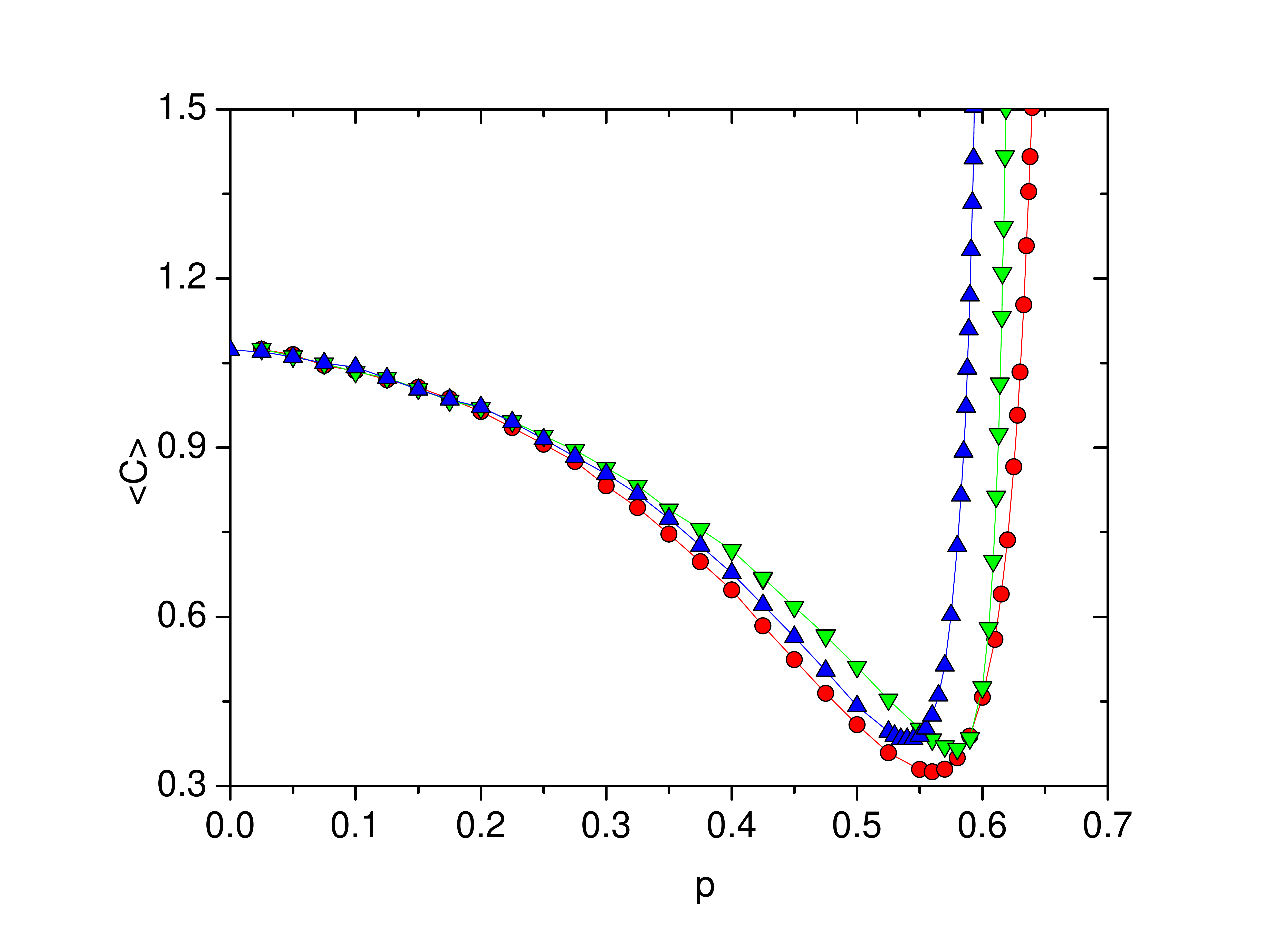}
\caption{(Color online)  Mean computational cost $\langle C \rangle $  as function of the imitation probability  $p$ for the  hierarchical network 
illustrated in Fig.\ \ref{fig:01}  $(\CIRCLE)$,  the scale-free network $(\blacktriangle)$, and  the random network $(\blacktriangledown)$.
The number of nodes is $L=125$ and the total number of links is $394$ for the three topologies.  
The parameters of the   rugged NK landscape are $N=16$ and $K=5$.  }
\label{fig:04}
\end{figure}

Figure  \ref{fig:04}  shows the mean computational cost for the hierarchical, small-world and random  networks in the case that 
all agents have the same imitation propensity $p$.  We find that  $\langle C \rangle $ is quite insensitive to variations on the topology of the network for small values of $p$. In particular, for $p=0$ one recovers the results of the independent search,  $\langle C \rangle \approx 1.08$, regardless of the topology. The topology becomes relevant only when the imitation propensity is large enough (say, $p > 0.55$) to allow the model string to drive the system  towards the local maxima and, in this case, the sensitivity to the topology is extreme.
This figure reveals that for large $p$ the system   can  easily  be trapped by the local  maxima,  from which escape can be extremely   costly.  This is akin to the groupthink phenomenon \cite{Janis_82},  when everyone in a group starts thinking alike,  which can occur when people put unlimited faith in a talented leader (the model strings, in our case). A similar maladaptive behavior induced by imitation (or, more generally, social learning)  has been observed in  groups of guppies \cite{Laland_98,Laland_11}. 

Most remarkably, Fig.\  \ref{fig:04}  reveals  that the hierarchical network consistently outperforms the other two topologies regardless of the value of the imitation propensity of the agents. It is expected that the presence of large hubs (or super-spreaders) will enhance
the performance of the system provided the information they broadcast is accurate \cite{Francisco_16}. This is not the case when the model
string and its followers are stuck in the neighborhood of a local maxima and this is the reason that the random network performs better than the scale-free network for $p>0.55$. Although the hierarchical network exhibits a super-spreader with a degree much larger than the hubs of the
scale-free network (see Figs.\ \ref{fig:01} and \ref{fig:02}), its modular structure somehow slows down the spreading of the inaccurate information through the system.

\begin{figure}[!h]
\centering\includegraphics[width=0.9\linewidth]{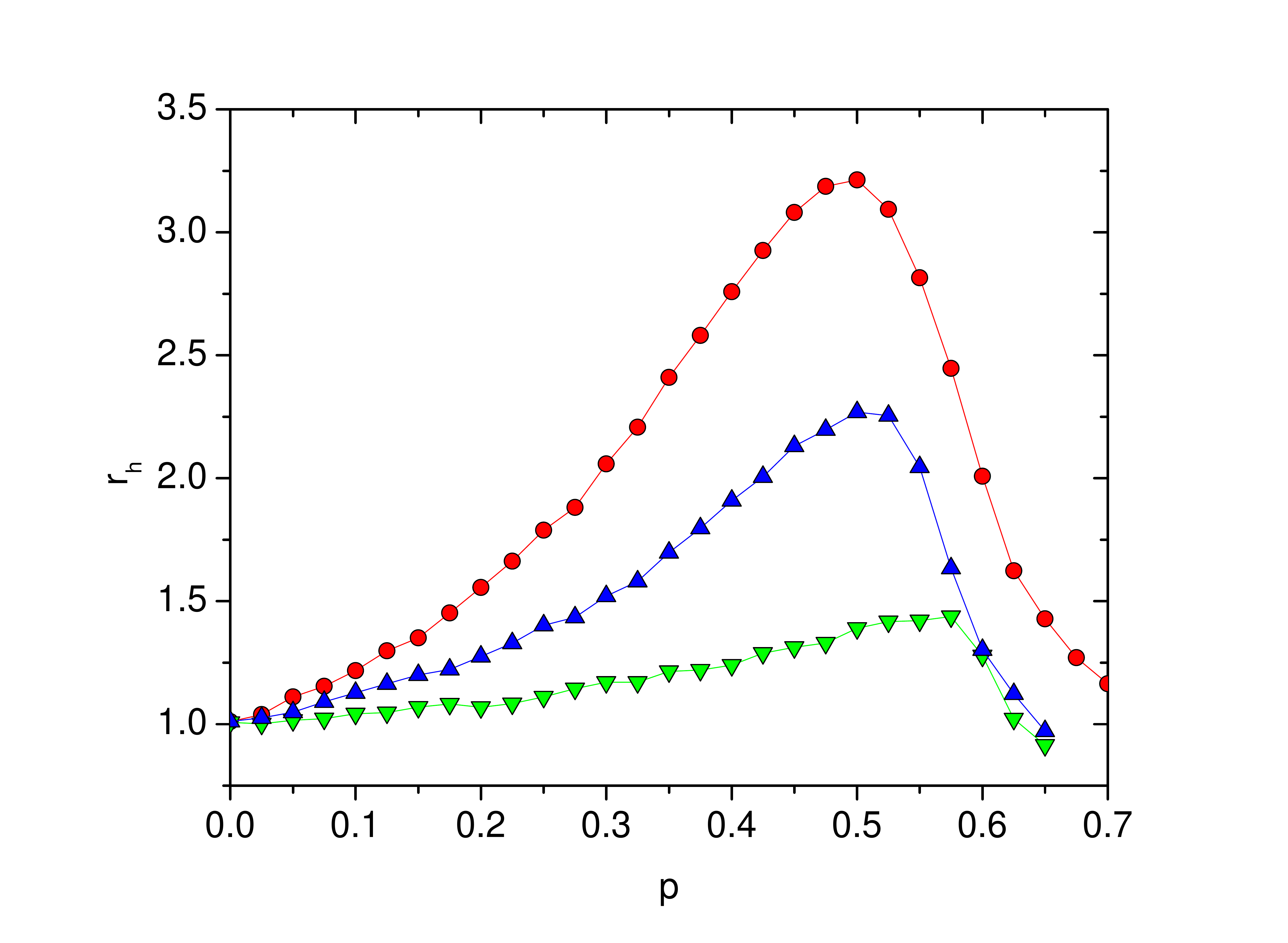}
\caption{(Color online) Ratio $r_h$ between the probability $P_h$  that the agent with the highest degree in the network finds the global maximum and the baseline probability $1/L$ for the case the $L$ agents are equiprobable to find that maximum. The symbols  
 $(\CIRCLE)$ are the results for the  hierarchical network,  $(\blacktriangle)$ are for the scale-free network, and $(\blacktriangledown)$ are for the random network $(\blacktriangledown)$.  The number of nodes is $L=125$ and the total number of links is $394$ for the three network topologies. 
The parameters of the   rugged NK landscape are $N=16$ and $K=5$.  }
\label{fig:45}
\end{figure}

It is instructive to examine  whether the degree of a node has any  influence on the chances that  the node finds the global maximum of the 
NK landscape.  To investigate this issue we  evaluate the probability $P_h$ that the  agent with the highest degree (main hub) is the one that finds the global maximum.   Since in the case all agents have the same probability of finding the global maximum  this probability is $1/L$, it is convenient to consider the  ratio  $r_h = P_h/\left ( 1/L \right ) = L P_h $ which  gives a measure of the  odds of the main hub to find the solution relative to a  situation where all agents are equiprobable to find the solution. Such  equalitarian situation occurs most notably in the independent search
(i.e., $p_i = 0, \forall i$) or in a regular lattice where all nodes have the same degree and imitation propensity.
In  Fig.\ \ref{fig:45} we show  $r_h$ as function of the imitation propensity $p$.  The results are qualitatively the same for  the three topologies under examination:  for small $p$ the degree of a node is of little relevance, as expected, and as $p$ increases and reaches
the optimal value at which the computational cost is minimum (see Fig.\ \ref{fig:04}), the importance of the degree of the node on its chances of finding the solution (and the consequent boosting of the overall performance of the system) is greatly amplified.  For large  $p$, however, the main hub seems to play a very detrimental role on the system performance. In fact,  its meager chances of finding the solution indicates that it is typically  trapped in a local maximum and, due to its large influence on the other agents, has dragged them together.

\begin{figure}[!h]
\centering\includegraphics[width=0.9\linewidth]{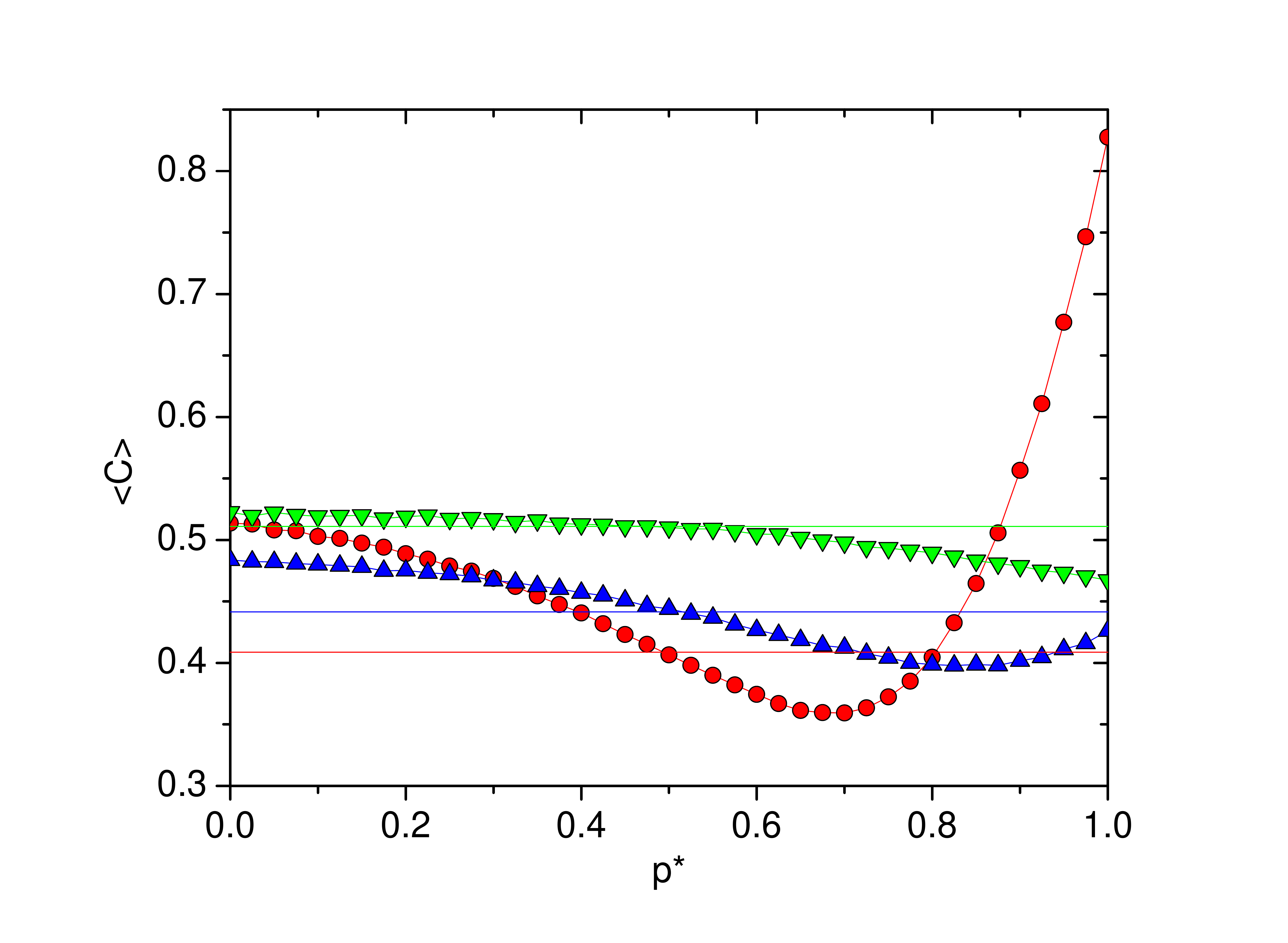}
\caption{(Color online)  Mean computational cost $\langle C \rangle $  as function of the imitation propensity  $p^*$ of the agent with the highest degree  for the  hierarchical network   $(\CIRCLE)$,  the scale-free network $(\blacktriangle)$, and the random network $(\blacktriangledown)$.
All other agents have their imitation propensities set  to $p=0.5$. The horizontal lines indicate the computational costs for  $p^* =p =0.5$.
The number of nodes is $L=125$ and the total number of links is $394$ for the three network topologies.  
The parameters of the   rugged NK landscape are $N=16$ and $K=5$.  }
\label{fig:05}
\end{figure}

 Our next experiment consists of picking the node with the highest degree and setting its imitation propensity  to the value $p^*$. The imitation propensities of the other $L-1$  agents in the system  are assigned  the same value $p$. In the case there are two or more nodes with the highest degree,  we assign the distinctive imitation propensity value  to only one of them. In this way we can examine the effect  of   disrupting the main  hub of the network  on the  problem-solving efficiency of the system .  The results for
 the three topologies are summarized in Fig.\ \ref{fig:05} for $p=0.5$
 
\begin{figure}[!h]
\centering\includegraphics[width=0.9\linewidth]{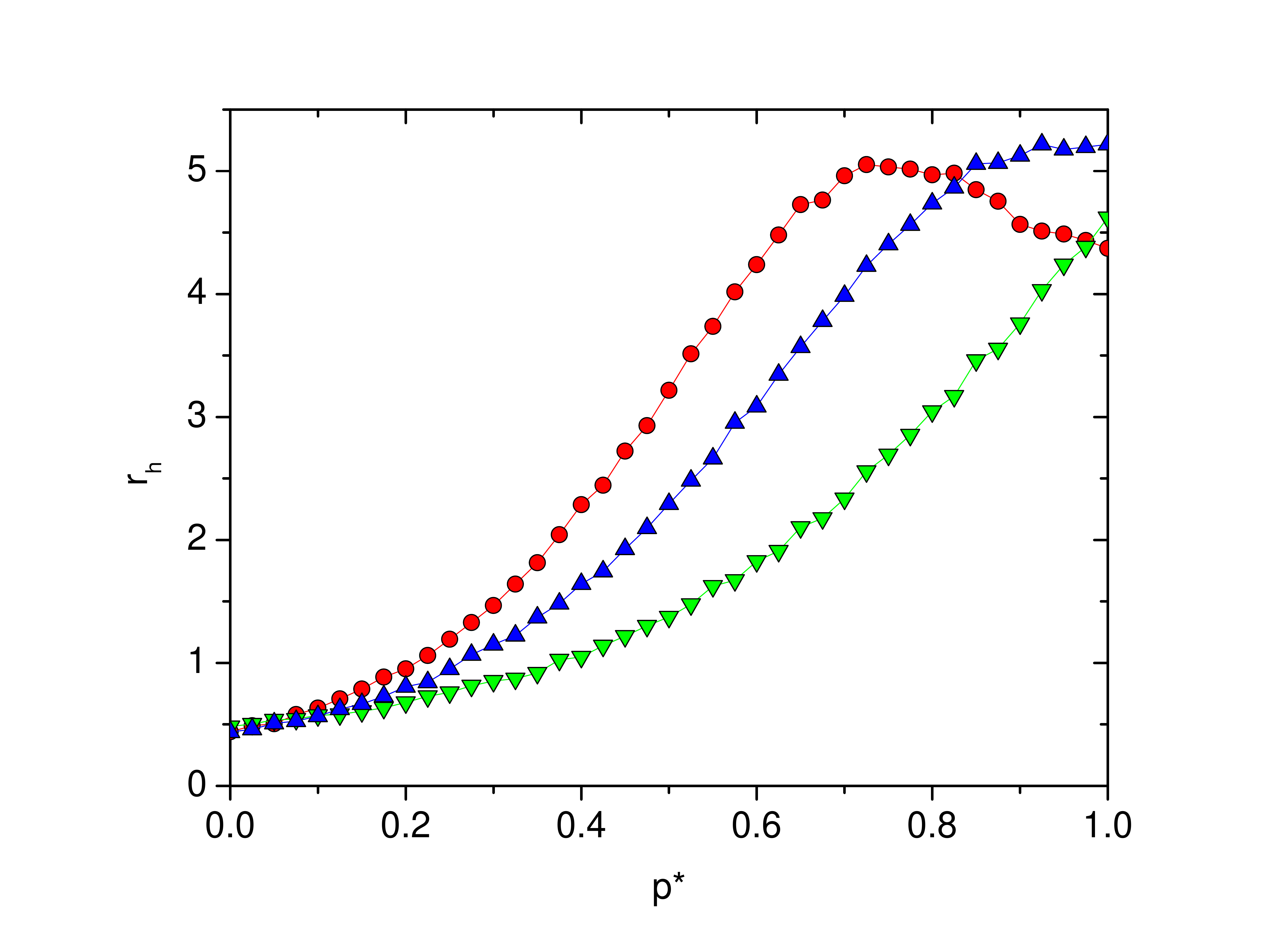}
\caption{(Color online) Ratio $r_h$ between the probability $P_h$  that the agent with the highest degree in the network, which 
has imitation propensity $p^*$,  finds the global maximum and the baseline probability $1/L$ for the case the $L$ agents are equiprobable to find that maximum. All other agents have imitation propensity set  to $p=0.5$. The symbols  
 $(\CIRCLE)$ are the results for the  hierarchical network,  $(\blacktriangle)$ are for the scale-free network, and $(\blacktriangledown)$ are for the random network $(\blacktriangledown)$. 
The parameters of the   rugged NK landscape are $N=16$ and $K=5$.  }
\label{fig:06}
\end{figure}
 
 As expected, the random network is the topology  less sensitive to perturbations in a hub, because its hubs have degrees that are not significantly larger than the degree of a typical node. It is interesting that whereas the performance of the random network is unaffected 
 for $p^* < p = 0.5$, it shows a slight  improvement  for $ p^* > p = 0.5$, which means that the presence of a compulsive imitator in the group can be advantageous, provided its influence on the rest of the group is limited. As for the more structured networks,  decreasing the imitation propensity of a hub relative to the rest of the group always results   in a decrease of  performance, whereas a moderate increase of the hub's relative imitation propensity can be highly  beneficial, though  compulsive imitation can lead to a  disastrous performance. These effects are much more pronounced in the hierarchical topology  because of the very high degree of its main hub. In fact, when  the second highest degree node is perturbed as well,  the scale-free network exhibits a  performance degradation for large $p^*$  similar to that observed for the hierarchical network   (data not  shown).
 
 We note that the computational cost is minimized for $p^* > p$ (see Fig.\ \ref{fig:05}), which suggests that  variability in the imitation propensities  of the agents may improve the performance of the cooperative system. This finding does not conflict with the claim that in a fully connected network  the optimal performance is achieved by a  homogeneous system \cite{Fontanari_16} because in the present analysis the agents differ in their degrees and so the situation $p=p^*$ does not correspond to a strictly homogeneous system.

Figure \ref{fig:06} shows the (relative) probability that the node with the  highest degree  finds the solution in comparison with  the equiprobable scenario
for the experiment summarized in  Fig.\ \ref{fig:05}. Since for $p^* \approx 0$ the main hub  executes an independent search,  its chances of finding the global  maximum are meager as it does not benefit from the experience of the other agents,  as expected.  Note, however, that the agents connected to that hub may imitate it with probability $p=0.5$, in case it happens to become the model string of their influence networks.  The  odds the main hub hits the solution increases monotonically with increasing  $p^*$, except for the hierarchical topology for which  $r_h$ reaches a maximum exactly at the value of $p^*$ that minimizes the computational cost (see Fig.\ \ref{fig:05}). The  subsequent  decrease of $r_h$ with increasing $p^*$ for $p^* \approx 1$ (compulsive imitation) as well as the disastrous performance of the system in this regime, highlight the nontrivial tradeoff between centrality and imitation propensity in the hierarchical network. 

\begin{figure}
\centering
\includegraphics[width=0.9\linewidth]{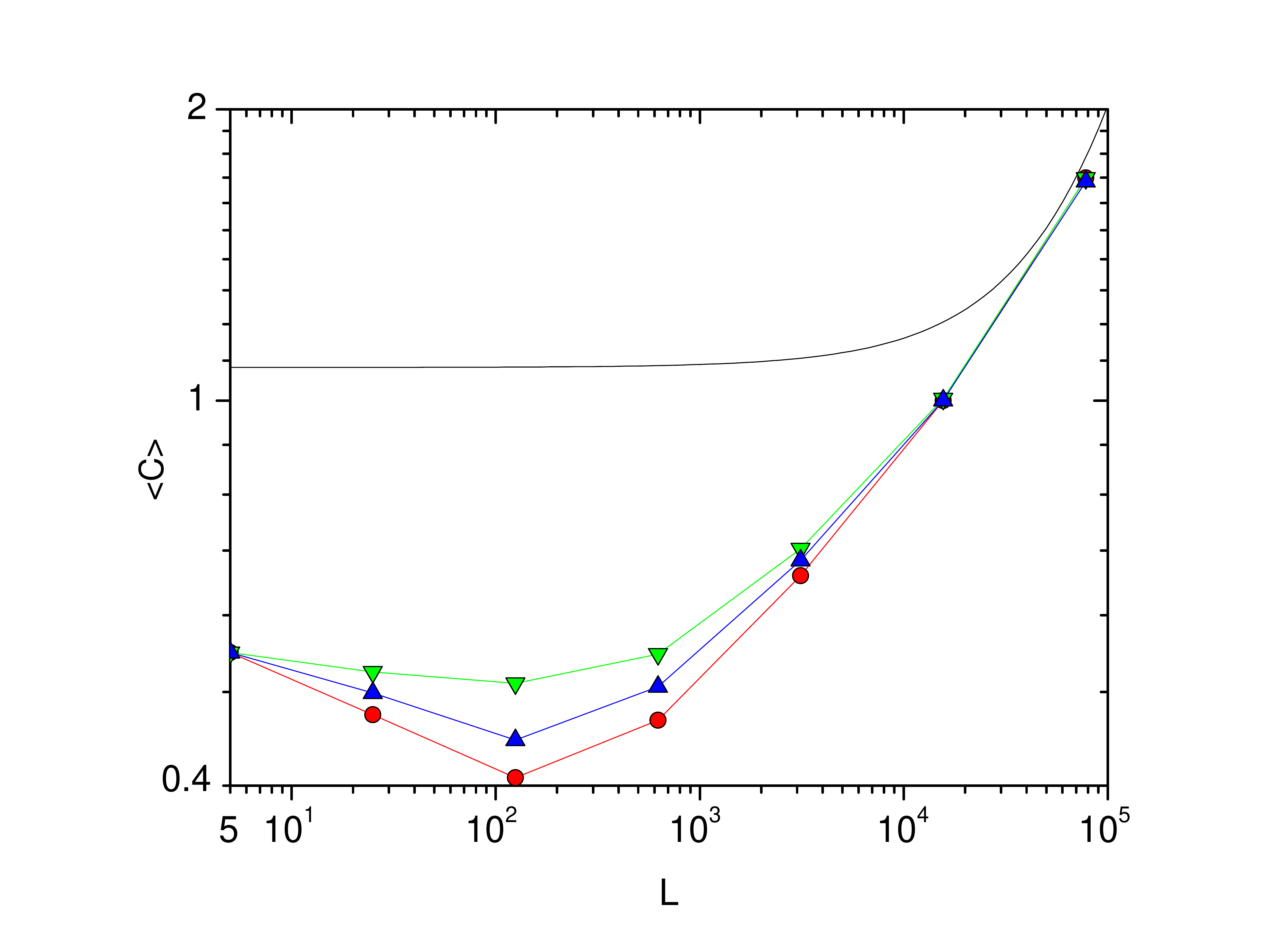}
\caption{(Color online) Mean computational cost $\langle C \rangle $   as function of the number of sites $L$ for the  hierarchical network   $(\CIRCLE)$,  the scale-free network $(\blacktriangle)$, and the random network $(\blacktriangledown)$. The 
imitation propensity  is $p=0.5$ for all agents.
The solid  curve is  the result for the independent search given by Eq.\ (\ref{Cind}).
 The parameters of the rugged NK landscape are $N=16$ and $K=5$. }
\label{fig:07}
\end{figure}
%

Finally, to conclude the analysis of the performance of the imitative search on a rugged landscape we address briefly two issues, namely,
the effects of changing the system size $L$ and the realization of the landscape. In Fig.\ \ref{fig:07} we show the  computational cost  as function of the  system size for the three network topologies and for the case that all agents have the same imitation propensity $p=0.5$. The total number of links is the same for all networks and it is determined by the number of links of the hierarchical network, namely, $ 6 \left( 5^{n-1} \right)+ 4 \left( 5^{n} - 4^{n} \right)$ where
$n = \ln L/\ln 5$.  This implies that for $L=5$ the network is fully connected (i.e., it has 10 links), regardless of the topology. As observed in previous analyses of the imitative search \cite{Fontanari_14,Fontanari_15}, for the three topologies  there is a system size   at which  the computational cost
is minimum. Provided that  the efficiency in solving problems has a selection value to the group members,  this finding may offer an alternative explanation for the size of groups of gregarious animals,  in addition to the more traditional selection pressures such as defense against predation, foraging success and the  managing of the social relationships \cite{Wilson_75,Kurvers_14,Dunbar_92}. We note  that, again, 
the hierarchical network outperforms the other two topologies, indicating that the combination of  the modularity and the scale-free 
properties produces a very efficient organization for distributed cooperative problem-solving systems.

\begin{figure}
\centering
\includegraphics[width=0.9\linewidth]{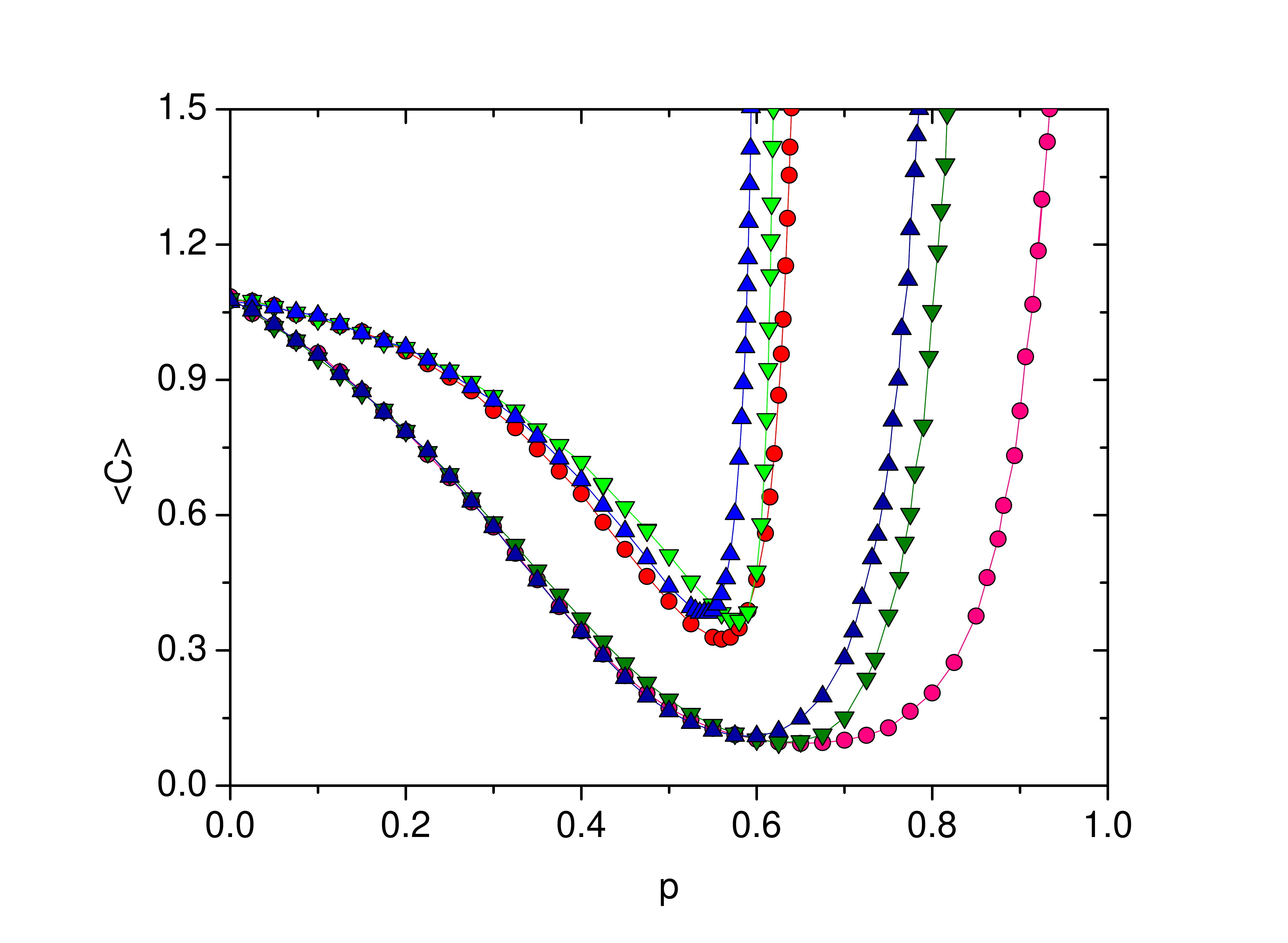}
\caption{(Color online) Mean computational cost $\langle C \rangle $  as function of   the imitation propensity $p$ 
for two realizations (the two bundles of data) of the NK fitness landscape  with $N=16$ and $K=5$. The symbols  
 $(\CIRCLE)$ are the results for the  hierarchical network,  $(\blacktriangle)$ are for the scale-free network, and $(\blacktriangledown)$ are for the random network.   }
\label{fig:75}
\end{figure}

To check the influence of the specific realization of the NK rugged fitness landscape we used in our study,   we have considered  several  random realizations of the landscape with $N=16$ and $K=5$.  In Fig.\ \ref{fig:75} we show the results for the original realization (see
Fig.\ \ref{fig:04}) and for a particular realization that produced a very different quantitative outcome. These quantitative differences are due to variations in the number of local maxima  for the different landscape realizations.
We find that, despite the quantitative differences, the dependence of  the computational cost on the imitation propensity  is the same for all landscape realizations (not only for the two realizations shown in the figure), namely, an initial  decrease towards an optimal value of  $p$, followed by a steep increase due to the trapping  in the local maxima. More importantly, the relative performances of the three topologies in the regime where the effect of the local maxima is critical are not altered by the landscape realization. In particular, the hierarchical  network outperforms the
other two topologies for all the realizations  that we have generated.
 
%

\begin{figure}[!h]
\centering\includegraphics[width=0.9\linewidth]{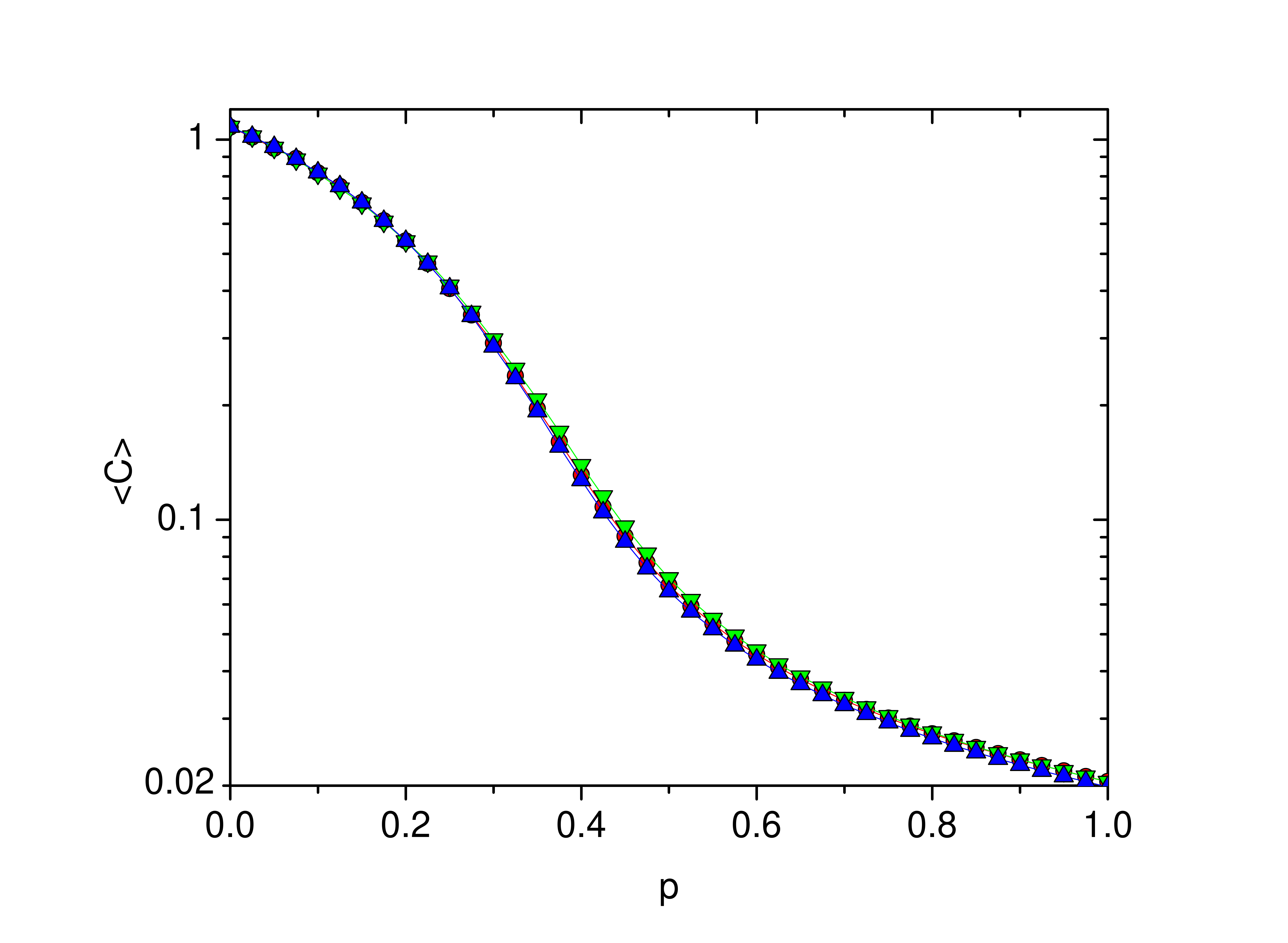}
\caption{(Color online)  Mean computational cost $\langle C \rangle $  as function of the imitation propensity  $p$ for the  hierarchical network 
illustrated in Fig.\ \ref{fig:01}  $(\CIRCLE)$,  the scale-free network $(\blacktriangle)$, and  the random network $(\blacktriangledown)$.
The number of nodes is $L=125$ and the total number of links is $394$ for the three network topologies.  
The parameters of the  smooth NK landscape are $N=16$ and $K=0$.  }
\label{fig:08}
\end{figure}

\subsection{Smooth landscape}

\begin{figure}[!h]
\centering\includegraphics[width=0.9\linewidth]{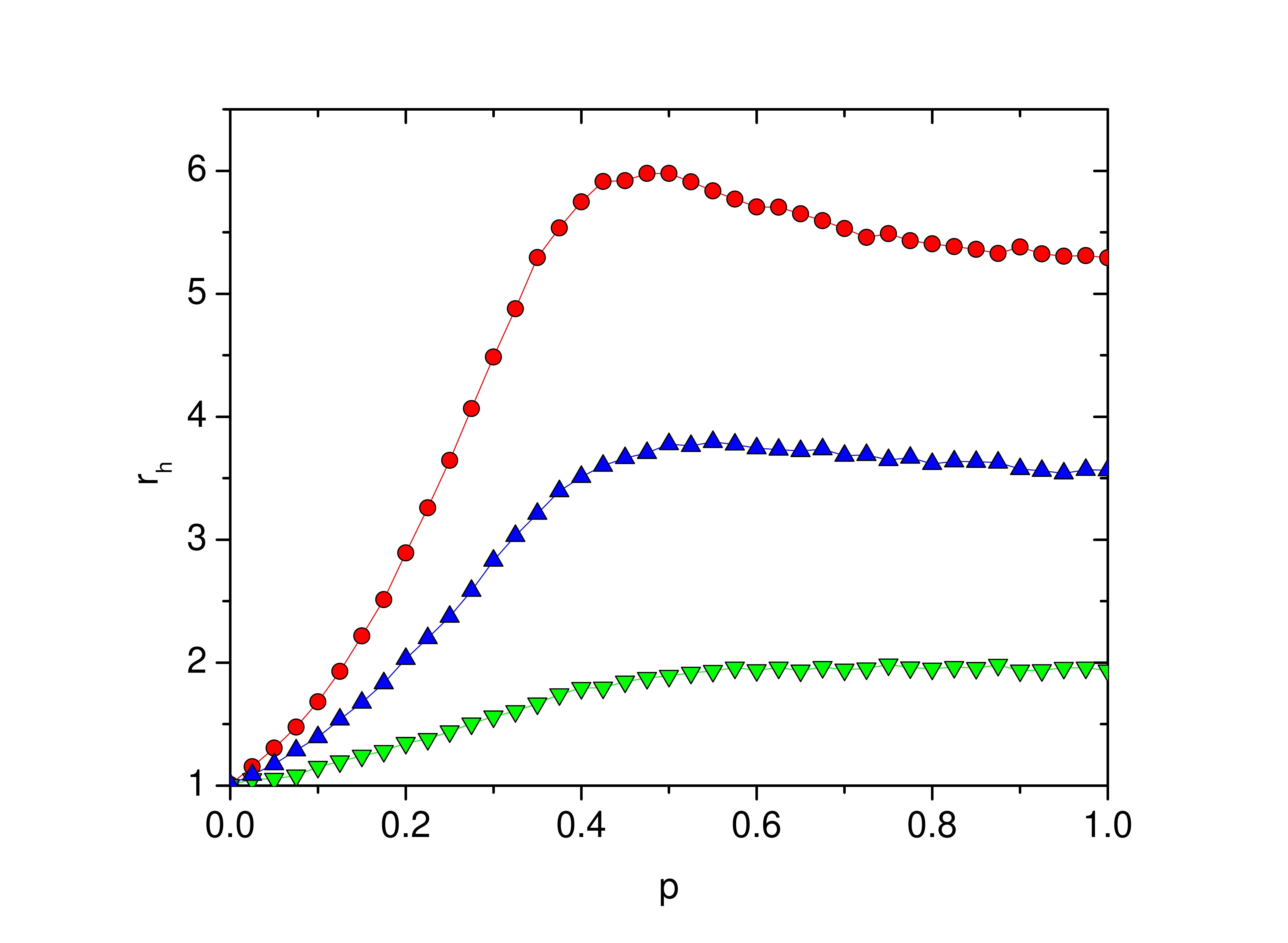}
\caption{(Color online) Ratio $r_h$ between the probability $P_h$  that the agent with the highest degree in the network finds the global maximum and the baseline probability $1/L$ for the case the $L$ agents are equiprobable to find that maximum. The symbols  
 $(\CIRCLE)$ are the results for the  hierarchical network,  $(\blacktriangle)$ are for the scale-free network, and $(\blacktriangledown)$ are for the random network $(\blacktriangledown)$. 
The number of nodes is $L=125$ and the total number of links is $394$ for the three network topologies.  
The parameters of the  smooth NK landscape are $N=16$ and $K=0$.  }
\label{fig:09}
\end{figure}

We turn now to the analysis of the performance of the imitative search on a smooth landscape  with $N=16$ and $K=0$. 
We note that for smooth landscapes ($K=0$) all landscape realizations are equivalent. 
Figure
\ref{fig:08} shows that in the absence of local maxima  the performances of  the  three topologies in the case  $p_i = p,  \forall i$ are practically indistinguishable in the scale of the figure and that the mean computational cost is a decreasing function of  $p$, i.e., the best performance is attained by always  imitating the model string ($p=1$)  and allowing only their clones to explore the landscape through the elementary move. Figure \ref{fig:09} shows the (relative) odds  that the main hub finds the maximum of the smooth landscape.  There is a strong correlation between the degree of a node and its chances of finding the maximum. The saturation of $r_h$ with increasing $p>0.5$ indicates that system has lost diversity -- all strings are clones or close neighbors  of the model string --
and so the odds of finding the solution is solely determined   by the degree of the node.    

\begin{figure}[!h]
\centering\includegraphics[width=0.9\linewidth]{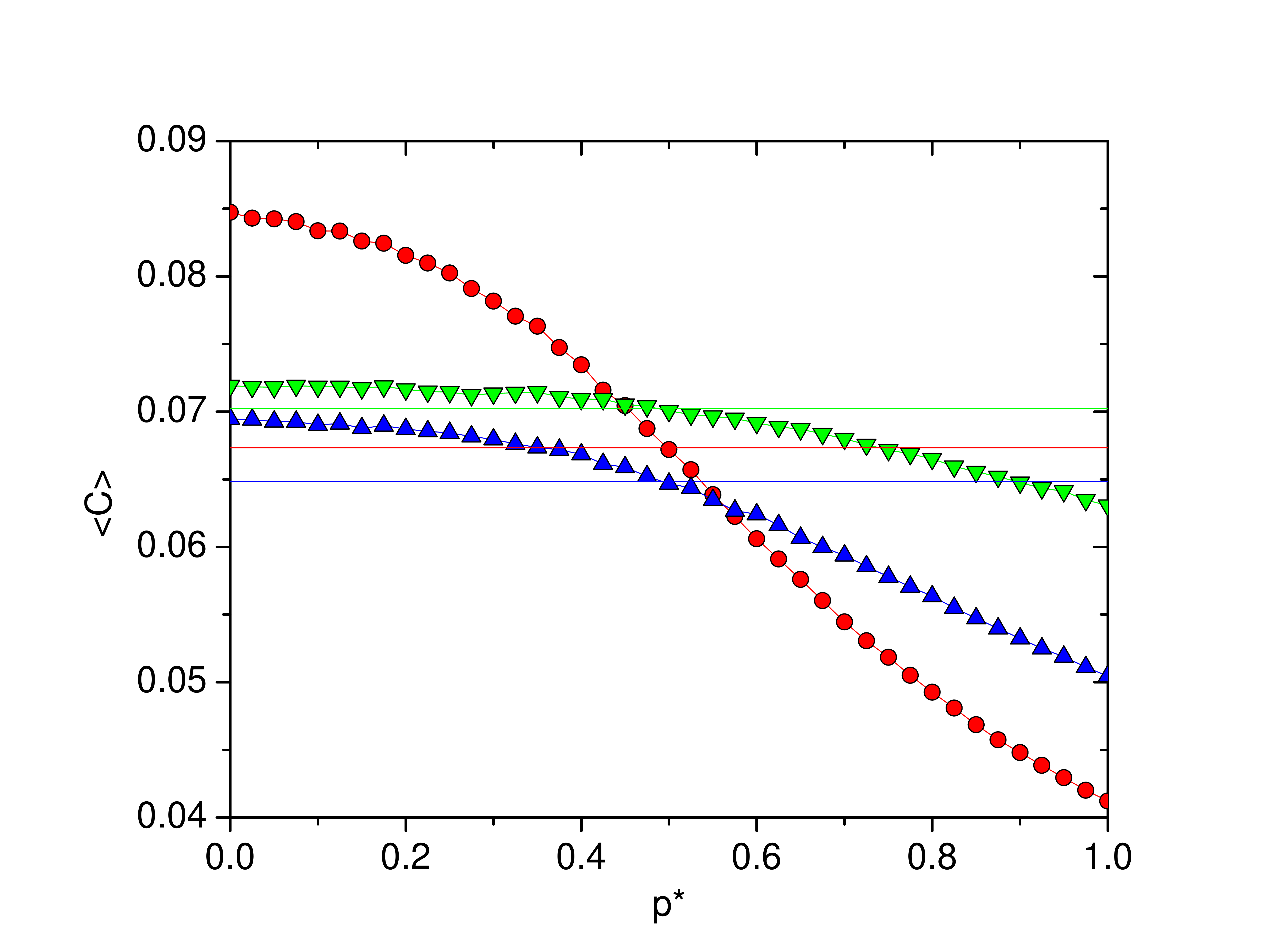}
\caption{(Color online) Mean computational cost $\langle C \rangle $  as function of the imitation probability  $p^*$ of the agent with the highest degree  for the  hierarchical network   $(\CIRCLE)$,  the scale-free network $(\blacktriangle)$, and the random network $(\blacktriangledown)$.
All other agents have imitation propensity set  to $p=0.5$. The horizontal lines indicate the computational costs of the homogeneous system $p^* =p =0.5$.
The number of nodes is $L=125$ and the total number of links is $394$ for the three network topologies.  
The parameters of the  smooth NK landscape are $N=16$ and $K=0$.  }
\label{fig:10}
\end{figure}

Although the performances of the three topologies are very similar  when all agents exhibit the same imitation propensities (see Fig.\ \ref{fig:08}), the situation changes remarkably when the main hub is assigned the differential imitation propensity $p^*$ as shown in Fig.\ \ref{fig:10}. 
Actually, in the scale of this figure we can observe that those performances in the  case $p^*=p =0.5$ are not identical and that
the scale-free network outperforms slightly  the other two topologies.
As in the case of the rugged landscape (see Fig.\ \ref{fig:05}), the performance of the random network is affected little by making its main hub explore the landscape independently of the other agents. The   performance of the hierarchical network, however,  is extremely sensitive to the influence of its main hub. For the smooth landscape, the monotone decreasing of the computational cost with increasing $p^*$ for all topologies is simply a consequence of the fact that copying the fittest string at the trial is always a certain step towards the solution of the problem.  This explains also the finding that the chances  that the main hub finds the maximum is a steep  increasing function of $p^*$ (data not shown).

\section{Discussion}\label{sec:disc}

Modularity is  ubiquitous  among biological entities, particularly  among processes and structures that can be modelled as networks 
\cite{Carroll_01,Lipson_07,Wagner_07}. A network is said to be  modular if it exhibits highly connected clusters of nodes that are scantily connected to nodes in other clusters. Modular systems are more adaptable since they are much easier to rewire  and be co-opted for another
functions than  monolithic networks. In addition, modular systems minimize the cost of the physical connections between nodes by  favoring short links and reducing long links \cite{Clune_13}. 

In this contribution we offer an extensive comparison between the performances of  distributed cooperative prob\-lem-solving systems that differ solely  by the topology of the network --  hierarchical, scale-free  and random --  that connects the agents in the system.  The number of nodes as well as the number of links are the same for all topologies examined.
The main difference between the hierarchical and the scale-free networks is the presence of  modular structures in the former topology (see Figs.\ \ref{fig:01} and \ref{fig:02}).
We show  that the hierarchical network  performs better than the other topologies  for imitative searches on rugged landscapes (see Fig.\   \ref{fig:07}). Since in this case the information broadcasted by the model strings (leaders) about the  location of the global maximum may be misleading  due to the presence of local maxima, the good performance of the hierarchical network  is really surprising because it has  a super-spreader that influences a vast number of agents in the system (see Fig.\ \ref{fig:01}). In fact,  in a previous study for the star topology,
it was shown that the presence of  a node with a very large degree facilitates the trapping of the system by the local maxima \cite{Francisco_16}. However, the modular structure of the hierarchical network somehow slows down the spreading of the inaccurate information through the system, resulting in a superior performance   compared with the other topologies.

In addition, we find that the hierarchical network is very sensitive to changes in the imitation propensity of its main hub (see Fig.\ \ref{fig:05}).  Interestingly, regardless of the topology and of the difficulty of the task, allowing the main hub to explore the landscape without much consideration for the other agents, even though those agents may learn from  it, is always detrimental to the performance of the system.  Except for the random network,  the optimal performance in a rugged landscape is achieved by letting the main hub to be a bit more propense to imitate its peers than vice versa. However, a compulsive imitator located at  the main hub of the hierarchical network  leads to a disastrous performance.
For the random network, where the main hub is not very influential, the performance is maximized by the compulsive imitation strategy. 
This conclusion holds true  for the three topologies  in the case of a smooth landscape (see Fig.\ \ref{fig:10}), but the reason is that in the absence of local maxima  it is always better to imitate  the fittest string in the group.

Since finding the global maxima of NK landscapes  with $K>0$ is an NP-Complete problem \cite{Solow_00}, one should not expect that the imitative  search (or, for that matter, any other search strategy)   would find those maxima much more rapidly
than the independent search, for which $\langle C \rangle \approx 1.08$ for $N=16$ and not too large system sizes. However, finding the solution much more slowly than the independent  search, as observed for large values of the imitation propensity $p$ (see Fig. \ref{fig:07}), is
a somewhat vexing outcome for any search strategy. But this negative outcome is akin to a maladaptive behavior associated to  social learning that has actually been observed in humans -- the Groupthink phenomenon \cite{Janis_82} -- and in  guppies \cite{Laland_98}. In this case,  a small founder group of guppies were trained to take an energetically costly circuitous route to a feeder and subsequently the  trained  members were gradually replaced by naive fishes. The experimenters found that even after 5 days in the tank, fish with founders trained to take the long route take the short route less frequently than unswayed fish.   This finding shows that maladaptive information can be socially transmitted through animal populations and it can hinder the learning of  the optimal behavior pattern
\cite{Laland_98,Laland_11}.

Our study of distributed cooperative prob\-lem-solving system deviates from the vast literature on cooperation that followed Robert Axelrod's 1984  seminal book The Evolution of Cooperation  \cite{Axelrod_84} since in that game theoretical framework it is usually assumed a priori that  mutual cooperation is the most rewarding strategy for the group. Here we consider a specific cooperation mechanism (imitation) and show that cooperation is not always beneficial, particularly in the case that the imitation propensity of the agents is large.  Since this happens because of  the misleading information being broadcasted by the model strings trapped by local maxima, an  efficient strategy to bypass this hindrance is to reduce the influence of the model string by decreasing the connectivity of the network \cite{Francisco_16}.  Interestingly, the finding that too frequent  interactions between agents harm the performance of the group (see Fig.\ \ref{fig:07}) may offer a theoretical justification for  Henry Ford's factory design in which the communication between workers was minimized in order to maximize productivity \cite{Watts_06} as well as for the scanty communication between leafcutters while they harvest leaves   \cite{Moffett_11}. Hence our conjecture that  the efficacy of imitative learning could impact on the  organization of  groups of animals capable of social learning.

\acknowledgments
The research of JFF was  supported in part by grant
15/21689-2, S\~ao Paulo Research Foundation
(FAPESP) and by grant 303979/2013-5, Conselho Nacional de Desenvolvimento 
Cient\'{\i}\-fi\-co e Tecnol\'ogico (CNPq).  SMR  was supported by grant  	15/17277-0, S\~ao Paulo Research Foundation (FAPESP).

\end{document}